\documentclass[preprint,authoryear,11pt,3p]{elsarticle}

\usepackage{setspace,mathtools,amssymb,siunitx,booktabs,kpfonts,xcolor}
\usepackage[font=itshape]{quoting}
\usepackage{pifont}

\usepackage{amssymb,multirow,amsmath,graphicx,graphics,array,lscape,rotating,color,amsthm,ifsym, colortbl, courier,listings,enumitem,hyperref,wrapfig}
\definecolor{dkgreen}{rgb}{0,0.6,0}
\definecolor{lightgreen}{rgb}{0.7,0.95,0.7}
\definecolor{gray}{rgb}{0.9,0.9,0.9}
\definecolor{mauve}{rgb}{0.58,0,0.82}
\definecolor{lightred}{rgb}{0.95,0.8,0.8}
\definecolor{lblue}{rgb}{0.2,0.45,0.6}
\lstdefinestyle{mystyle}{frame=tb,
  language=R,
  aboveskip=3mm,
  belowskip=3mm,
  showstringspaces=false,
  columns=flexible,
  basicstyle={\footnotesize\ttfamily},
  numbers=none,
  numberstyle=\tiny\color{gray},
  keywordstyle=\color{blue},
  commentstyle=\color{dkgreen},
  stringstyle=\color{mauve},
  breaklines=true,
  breakatwhitespace=true,
  tabsize=3
}
\usepackage{bbm}
\usepackage{makecell}
\usepackage{fancyhdr}

\begin{document}

\fancypagestyle{pprintTitle}{%
\lhead{TEXT} \chead{}\rhead{\scriptsize \thepage}
\lfoot{}\cfoot{}\rfoot{}
\renewcommand{\headrulewidth}{0.0pt}
}

\begin{frontmatter}

\title{Wielding Occam's razor: Fast and frugal retail forecasting}

\author[add1]{Fotios Petropoulos}
\ead{f.petropoulos@bath.ac.uk}
\author[add2]{Yael Grushka-Cockayne}
\ead{grushkay@darden.virginia.edu}
\author[add3]{Enno Siemsen}
\ead{esiemsen@wisc.edu}
\author[add4]{Evangelos Spiliotis}
\ead{spiliotis@fsu.gr}

\address[add1]{School of Management, University of Bath, UK}
\address[add2]{Darden School of Business, University of Virginia, USA}
\address[add3]{Wisconsin School of Business, University of Wisconsin-Madison, USA}
\address[add4]{Forecasting and Strategy Unit, School of Electrical and Computer Engineering, \\National Technical University of Athens, Greece}

\begin{abstract}
The algorithms available for retail forecasting have increased in complexity. Newer methods, such as machine learning, are inherently complex. The more traditional families of forecasting models, such as exponential smoothing and autoregressive integrated moving averages, have expanded to contain multiple possible forms and forecasting proles. We question the complexity of forecasting and the need to consider such large families of models. Our argument is that parsimoniously identifying suitable subsets of models will not decrease forecasting accuracy, nor will they reduce the ability to estimate forecast uncertainty. We propose a framework that balances forecasting performance versus computational cost. As a result, we consider a reduced set of models. We empirically demonstrate that such a reduced set performs well. Finally, we translate computational benefits to monetary cost savings and environmental impact and discuss the implications of our results in the context of large retailers.
\end{abstract}


\end{frontmatter}
\thispagestyle{fancy}

\section{Introduction}\label{sec:introduction}

Retail demand forecasting is at a crossroads. Complex machine learning (ML) methods have become more prevalent and are pioneered by industry champions \citep{Seaman2018-xf}. Results from field tests and forecasting competitions demonstrate that using these methods can lead to forecast accuracy improvements \citep{Ma2021, Makridakis2020-wq}. Research on demand planning in retail has increased significantly over the past 15 years \citep{Rooderkerk2022}, and retail analytics is becoming an increasingly important research topic \citep{Caro2020} as long-held industry practices are overhauled through constant innovation. Examples of such innovations include the effective use of analogies via ML in new product forecasting for flash sales \citep{Ferreira2016}, the effective use of aggregation for prediction \citep{Cohen2022}, or the use of social media information in demand forecasting \citep{Cui2018}.

At the same time, complex methods can be opaque, and their forecasts are often difficult to explain to non-experts. Further, the forecasting task for modern omnichannel retailers is immense. For instance, Walmart features approximately 200 thousand unique stock-keeping units (SKUs) and more than 10 thousand stores, resulting in 2 billion unique SKU-store combinations \citep{Seaman2018-xf}. Each SKU may require a detailed weekly, or even daily forecast, depending on replenishment schedules. Online markets can require more than a trillion unique combinations of SKU-ZIP codes \citep{Seaman2018-xf}.\footnote{See also the respective Kaggle competition: https://www.kaggle.com/c/walmart-recruiting-store-sales-forecasting/data} The sheer scale involved can be costly for retailers, with best practices requiring running multiple models per forecast cycle to leverage combinations, fine-tuning models regularly, and generally running complex models over all their products and markets. The cost and environmental impact of the necessary CPU power can be immense. Faced with this complexity, should retailers - instead of relying on more and more complex methods - fall back on simple forecasting methods?

The number of series to forecast increases further if one considers that retail organizations require forecasts in hierarchical structures (i.e., store-level sales aggregating to regional sales) with decisions required at different levels of the hierarchy. Just as store-level forecasts are necessary for store replenishment, regional-level forecasts  are necessary to plan the replenishment of distribution centers. Although traditional bottom-up or top-down methods do not require separate forecasts at different levels, modern optimal reconciliation forecasts do \citep{Hyndman2011}.  \citet{Yelland2019} described how Target handles its forecast volume. Large retailers are not alone in their need for forecasting at scale. For example, Google needs product forecasts (such as ads, search, YouTube, Analytics, etc.) to plan its resources, infrastructure, and data center operations.

The challenge of increased computational cost becomes even more relevant when forecasters may face management or peer pressure to adopt leading-edge machine-learning approaches in their forecasting applications. The debate about whether ML-based approaches or simpler and faster statistical methods are more useful in forecasting is ongoing. \cite{Makridakis2018} showed that ML-based approaches fall short of statistical benchmarks when applied to each series separately. In the subsequent M4 and M5 forecasting competitions, however, ML-based approaches applied in a cross-sectional fashion did outperform statistical ones, but the performance differences were small \citep{Makridakis2019-oy,Makridakis2020-wq}.

Our paper studies anew the decades-old tension between complex and simple forecasting methods in large-scale forecasting tasks. When Box-Jenkins models appeared, statisticians hailed them for their superior performance; but forecasting competitions failed to establish their advantage over simple exponential smoothing methods \citep{Makridakis2000a}. In the aftermath, exponential smoothing itself became more complex in the context of the exponential smoothing (ETS) framework, which conceptually distinguished between 30 exponential smoothing models. With the advent of modern data science, machine learning was reintroduced and grew in importance. We argue it is time for reassessment. These more complex methods should only be implemented at scale if they have demonstrable performance advantages. From a forecast value-added perspective, their benefits should outweigh the downsides of increased computational complexity. This is especially true in retail organizations, where small gains in forecast accuracy do not always equal operational improvements \citep{Kolassa2021M5}.

Our main research question is as follows: Can we maintain the accuracy of statistical forecasts while reducing the cost required to generate them? Method complexity is an important concept to help answer that question. Higher complexity promises a better correspondence between the method and the underlying data-generating process, thus possibly increasing the accuracy of forecasts. At the same time, complexity usually means increased computational effort.

Although traditional time series methods are simpler than machine learning, they also rely on unnecessarily complex families of models. Our paper thus  examines how existing families of methods can be simplified; we develop reduced sets of models, and we compare such reduced sets to complete sets to establish that reducing the sets comes with little reduction in accuracy while enabling efficiency. We then proceed to compare these reduced sets of models to modern machine learning approaches in a large-scale retail data set to demonstrate that retailers could benefit from such a simplified forecasting approach by increasing their efficiency and reducing the environmental impact of their forecasting without sacrificing accuracy.

Complex models may not perform as well as hoped because of overfitting. The stochastic processes that generate real-world data change over time. Thus, identifying the "optimal" forecast in an existing data set might not be effective because the future will not look exactly like the past. Although some forecasting models produce very good in-sample fits, their out-of-sample performance is not as good.

\cite{Goldstein2009-rl} demonstrate that fast and frugal heuristics often used in human judgment can give remarkably precise estimates. These fast and frugal heuristics achieve their robustness through simplicity. The authors warn practitioners about the use of overly complex models: ``The  danger  is  that complex   methods   become   an   end   in themselves, a  ritual  to  impress  others,  and  at  the  same  time opportunities  to  learn  how  to  do  things  better  are missed. Learning requires some form of transparency, which forecasters  can  best  achieve  when  they understand what they are doing'' \citep[][p. 770]{Goldstein2009-rl}. We thus propose that fast and frugal forecasting may be beneficial for retail firms.

In related research, \cite{Nikolopoulos2018-wa} examined how a suboptimal search for the smoothing parameter in the simple exponential smoothing method will lead to an average forecasting performance that is statistically indifferent to the more computationally expensive, optimal search. The authors conclude that accepting suboptimal solutions will not affect the quality of the forecasts but will save significant resources in terms of producing these forecasts. 

We first focus on two well-studied and common forecasting families: exponential smoothing \citep{Hyndman2002} and Autoregressive Integrated Moving Average (ARIMA) \citep{box2008}. Automatic algorithms based on information criteria have been proposed to select the best model within each of these two families \citep{Hyndman2008b}. Fitting and parameterizing all applicable models, however, can be computationally expensive. 

We empirically explore the performance of subsets of models in terms of forecasting accuracy achieved and the computational resources required for both the exponential smoothing and the ARIMA families. We present the value added in each step of the reduction process and explore the differences in the selected model forms. For both model families, we show that significant reductions in cost can be achieved without harming - and in some cases even improving - forecasting performance. We then proceed to compare these reduced sets of models to modern machine learning techniques in the context of large-scale retail forecasting.


\section{Time series forecasting with exponential smoothing and ARIMA models}\label{sec:litreview}
In this section, we provide an overview of two families of time series models, highlighting the richness of options each offers and how model variants are selected. We then consider the need for more efficient reduced forms or subsets of such models. 

\subsection{The exponential smoothing family of models}\label{sec:ets}

Exponential smoothing models are univariate time-series forecasting models. They work on the assumption that newer data have greater relevance than older ones and, as such, should count more in calculating future point forecasts. The exponential smoothing family of models can handle level, trend, and seasonal patterns. Separate equations are used to exponentially smooth each of these components.

The Simple Exponential Smoothing (SES) method was proposed by \cite{Brown56} and could handle level-only data. This method was later extended by \cite{Holt20045} and \cite{Winters1960} to include trend and seasonal cases. However, assuming a constant trend in the exponential smoothing method results in significant and positively biased forecasts for longer horizons. Thus, \cite{Gardner1985b} proposed that trends should be dampened toward a flat line, emulating the mature stage of product life cycles. Damped-trend exponential smoothing is a very robust forecasting method that is hard to outperform. Forecasting researchers use it even today as an objective forecasting benchmark \citep{Gardner2006,Makridakis2020-eu,Makridakis2019-oy}. 

\cite{pegels1969} and \cite{Gardner2006} proposed a taxonomy for exponential smoothing models containing 12 forecasting profiles that are combinations of possible trend and seasonal components. They suggest that an observed trend might be nonexistent (no trend; level data), linear (additive), exponential (multiplicative), or damped. At the same time, data might be nonseasonal or have additive or multiplicative seasonality.

For years, forecasters considered exponential smoothing methods unable to capture the data generation process as well as more sophisticated approaches. \cite{Makridakis82} and \cite{Makridakis2000a}, however, have shown that exponential smoothing performed as well, if not better, than more complex methods. Given their performance, simplicity, and intuitive mechanics, it is unsurprising that these methods (SES, Holt's, Holt-Winter's, and damped-trend) rank among the most popular ways to forecast time series. They are implemented in most commercial forecasting support systems (e.g.  ForecastPro, SAP APO, and SAS) and are also frequently used by companies \citep{Fildes2009, Petropoulos2018-mt}.

One major theoretical development of exponential smoothing models was their revamp into a statistically sound state-space framework \citep{Ord1997-aq, Hyndman2002, Taylor2003715}. The introduction of this framework came with several advantages, such as easy estimation of the uncertainty of forecasts (through prediction intervals) and estimation of the likelihood of these models. The latter simplified the automatic model selection process across exponential smoothing methods by using information criteria (see Section \ref{sec:ics} for a formal description of information criteria). Finally, \cite{Hyndman2002} introduced models with both additive and multiplicative errors, increasing the forecast profiles of \cite{pegels1969} to 24. \cite{Taylor2003715} further extended these to 30, adding multiplicative damped-trend models.

\cite{Hyndman2008-iu} codified the 30 exponential smoothing models with an acronym for their three components: Error, Trend, and Seasonality (ETS). As such, Holt's exponential smoothing model can be expressed as AAN, meaning a model with additive error, additive trend, and no seasonality. Similarly, MAdM is a model with multiplicative error, damped additive trend, and multiplicative seasonality. The acronyms for all 30 models are given in Table \ref{tab:allmodelsets}.

Of these 30 models, 11 sometimes lead to estimation difficulties and infinite forecast variances for long  horizons \citep[see][Chapter 15]{Hyndman2008-iu}. For this reason, some modern forecasting software (such as the \textit{forecast} package for the R statistical software) by default avoid these models. The 19 remaining models highlighted in boldface in Table \ref{tab:allmodelsets} will be referred to as ``all applicable models'' or simply as ``all models.'' Four of these models (MMN, MMdN, MMM, and MMdM) lack analytical expressions, so simulation approaches are used to obtain prediction intervals. This significantly increases calculation times in practice.

\begin{table}[h]\small
\centering
\caption{Exponential smoothing models. Applicable models are highlighted in boldface.}\vspace{0.25cm}
\begin{tabular}{cccccccc}
\Xhline{2\arrayrulewidth}
\multicolumn{4}{c}{Additive \textbf{E}rror} & \multicolumn{4}{c}{Multiplicative \textbf{E}rror} \\
\cmidrule(lr){1-4}
\cmidrule(lr){5-8}
& \multicolumn{3}{c}{\textbf{S}easonality} & & \multicolumn{3}{c}{\textbf{S}easonality} \\
\cmidrule(lr){2-4}
\cmidrule(lr){6-8}
\textbf{T}rend & N & A & M &                      \textbf{T}rend & N & A & M \\
\Xhline{2\arrayrulewidth}
N &  \textbf{ANN} &  \textbf{ANA} &  ANM & N &  \textbf{MNN} &  \textbf{MNA} &  \textbf{MNM} \\
A &  \textbf{AAN} &  \textbf{AAA} &  AAM & A &  \textbf{MAN} &  \textbf{MAA} &  \textbf{MAM} \\
Ad &  \textbf{AAdN} &  \textbf{AAdA} &  AAdM & Ad &  \textbf{MAdN} &  \textbf{MAdA} &  \textbf{MAdM} \\
M &  AMN &  AMA &  AMM & M &  \textbf{MMN} &  MMA &  \textbf{MMM} \\
Md &  AMdN &  AMdA &  AMdM & Md &  \textbf{MMdN} &  MMdA &  \textbf{MMdM} \\
\Xhline{2\arrayrulewidth}
\end{tabular}
\label{tab:allmodelsets}
\end{table}

Extensions to exponential smoothing models include models with regressor variables \citep[also known as ETSx:][]{Hyndman2008-iu}, models with multiple seasonal patterns \citep{Taylor2003-ye,Taylor2008-qu}, models with Box-Cox transformation, autoregressive and moving-average errors, trend, and seasonal components \citep{De_Livera2011-tw}, a model for forecasting product life cycles \citep{Guo2018-jn}, and a complex exponential smoothing model \citep{Svetunkov2016-hz}. A model that performed exceptionally well in the M5 forecasting competition at the SKU level is based on exponential smoothing \citep{deRezende2022}.

\subsection{The ARIMA family of models}\label{sec:arima}

ARIMA models offer an alternative to exponential smoothing models. In contrast to the exponential smoothing family that consists of a finite set of 30 models, the ARIMA family consists of an infinite number of models \citep{box2008}. The basic mechanism of ARIMA models resembles that of multiple regression models, in which the predictors are either lagged values of the data (autoregressive, AR) or lagged values of the forecast errors (moving averages, MA). The regression modeling is applied to the differenced, stationary data.



A nonseasonal univariate ARIMA model is generally denoted as ARIMA$(p,d,q)$, where $p$ is the order of the autoregressive term, $q$ is the order of the moving average term, and $d$ is the degree of first differencing required to achieve stationarity. Seasonal ARIMA models add seasonal autoregressive and moving average terms while accounting for seasonal differences. A seasonal ARIMA model is denoted as ARIMA$(p,d,q)(P,D,Q)_{s}$ with $P$ and $Q$ as the seasonal autoregressive and moving average terms; $D$, the degree of seasonal differences; and $s$, the periodicity of the data. All $p$, $d$, $q$, $P$, $D$, and $Q$ are non-negative integers. For example, ARIMA$(0,1,2)(1,1,0)_{12}$ refers to an ARIMA model for monthly data ($s=12$) with first-order non-seasonal and seasonal differences ($d=D=1$), a second-order non-seasonal MA term ($q=2$), and a first-order seasonal AR term ($P=1$).

ARIMA models excel at handling a remarkable range of data generation processes. In fact, in the 1970s, ARIMA was considered the holy grail in forecasting, especially in  econometrics circles, because no other family of models could produce superior goodness-of-fits. However, this came with two drawbacks. First, the estimation of an appropriate ARIMA model given an observed series was a combination of art and science, detracting from their popularity in the industry and creating their reputation as "black boxes." Second, despite their good performance in-sample, ARIMA models did not perform as well in  out-of-sample forecasting performance \citep{Makridakis1979-lv,Makridakis82}; effectively, ARIMA models (and especially large ARIMA models) suffered from overfitting. The same insights were presented in a much later study of the results of the M3 forecasting competition \citep{Makridakis2000a}.

\cite{Hyndman2008b} offered an automatic algorithm to specify and parameterize an ARIMA model. The algorithm starts by applying repeated unit root tests\footnote{Kwiatkowski-Phillips-Schmidt-Shin (KPSS) test.} to determine the degree of first-order and seasonal-order differencing required to render a series stationary. Then a stepwise selection process is applied as follows. An initial set of models is estimated, and the best one is selected based on an information criterion. A new set of candidate models is then considered by varying the values of $p$, $q$, $P$, and $Q$ of the currently selected model by $\pm 1$ while also considering models with and without a constant. If a better model is found, then the search continues; otherwise, the search stops.

A non-stepwise search for an optimal ARIMA model is also possible. However, it is necessary to limit the maximum order ($p+q+P+Q$) of the search. Increasing the maximum order increases the search space for an optimal model form but also significantly increases the computational time required to complete the search. In this paper, we explicitly examine the balance between the simplicity of the ARIMA models (effectively the maximum order for ARIMA models, which directly reflects the computational complexity) and  their forecasting performance.

\subsection{Toward reduced families of forecasting models}\label{sec:reduced}

The above discussion focused on the development of two families of models: exponential smoothing and ARIMA. Especially with the former, we noticed an increase in the variety of models over the years. At the same time, the theory of ARIMA models, which was developed by Box and Jenkins in the 1970s, suggests an infinite number of possible models. Such large families of models are linked by an assumption that capturing as many forecast profiles as possible will yield improved forecasting performance. Moore's law and increases in computing power and speed enabled the use of richer and larger families of models. But are all of them really needed? Would our forecasts suffer if we restricted ourselves to small subsets of models within these families? Would model selection be easier and more efficient if we had to select among fewer models?

We extended the studies of \cite{Makridakis1983} and \cite{Makridakis2000a} to search for parsimony within each of these two widely-used families of models. In line with the simplicity argument \citep{Green2015-gf} and Occam's razor, we suggest that parsimony in forecasting should be preferred over complexity, especially if performance is maintained. Ultimately, we suggest that the cost of a forecast error (or forecast utility) should be balanced against the cost of generating the forecast. 

\section{Experimental design}\label{sec:design}

This section outlines an experiment in which we tested whether a reduced set of models performs well with real-world data. In our experiment, we considered 50,045 time series to compare the performance of a selected model from a given pool to the effort required to search for that model within the pool. We constructed five pools of exponential smoothing models and eight pools of ARIMA models. These exponential smoothing pools are described in  \ref{sec:pools_ets} and the ARIMA pools in \ref{sec:pools_arima}. We describe the selection procedure of each model from a considered pool in \ref{sec:ics}. The performance metrics used for measuring the performance of the selected model and the cost of searching within a pool are described in \ref{sec:performance}. Finally, the data used in this study is described in \ref{sec:data}. 

\subsection{Reduced pools of exponential smoothing models}\label{sec:pools_ets}

We explored the performance of different subsets of models within the exponential smoothing framework. In total, we examined the following five pools of exponential smoothing models:
\begin{itemize}[noitemsep,nolistsep]
\item All applicable exponential smoothing models within the \texttt{ets} function of the \textit{forecast} package for the R statistical software. This includes the 19 models (8 when data are nonseasonal) highlighted in boldface in Table \ref{tab:allmodelsets}. We will refer to this pool as ``all models''.
\item All applicable exponential smoothing models within the \texttt{ets} function, excluding those models with multiplicative trends. These models are excluded because, even if they are interesting from a theoretical point of view, they lead to unrealistic  forecasts.\footnote{The \texttt{ets()} function of the \textit{forecast} package for the R statistical software by default excludes models with multiplicative trends (since its 6.0 version, May 2015). However, their application is still possible through the \texttt{allow.multiplicative.trend} argument.} Additionally, as discussed in Section \ref{sec:ets}, these models use simulation to find prediction intervals and are computationally expensive.
This leads to a collection of 15 models (6 when data are nonseasonal), with four models excluded compared with the previous case (MMN, MMdN, MMM, MMdM). We will refer to this pool as ``no multiplicative trend.'' \cite{Hyndman2008-iu} mentioned that linear-only models might be indeed considered but did not investigate the performance of such subsets. 
\item All applicable exponential smoothing models within the \texttt{ets} function, excluding non-damped trended models when trend exists. Damped exponential smoothing has been found to significantly outperform Holt's linear trend exponential smoothing, which tends to produce positively biased forecasts. Moreover, damped exponential smoothing is used as one of the benchmarks in empirical forecasting evaluations, with \cite{Gardner2006} noting that ``it is still difficult to beat the application of a damped trend to every time series''. This leads to a collection of 12 models (five when data are nonseasonal), with seven models excluded compared with the first case (AAN, AAA, MAN, MAA, MAM, MMN, MMM). We will refer to this pool as a ``damped trend''.
\item All applicable exponential smoothing models within the \texttt{ets} function, excluding those in which the error and seasonality type do not match. Models with additive error and multiplicative seasonality could lead to numerical difficulties and have already been excluded. We also suggest excluding models with multiplicative error and additive seasonality because they tend to overestimate the variance and lead to very large prediction intervals. This leads to a collection of 16 models (eight when data are nonseasonal), with three models excluded compared with the first case (MNA, MAA, MAdA). We will refer to this pool as ``match error with seasonal type'.'
\item All applicable exponential smoothing models within the \texttt{ets} function, excluding those in the three above cases. This leads to a collection of eight models (four when data are nonseasonal). We will refer to this pool as ``reduced.'' 
\end{itemize}

It is noteworthy that, from the pools of models described above, only the last one (reduced exponential smoothing) offers a balance with regard to the number of models considered for each of the following four broad categories (profiles): level only, trend only, seasonal only, trend and seasonal; see Table \ref{tab:exppoolsmodelssummary}.

\begin{table}[h]\small
\centering
\caption{Exponential smoothing models in the four broad forecast profiles consider by each pool.}\vspace{0.25cm}
\begin{tabular}{lcccc}
\Xhline{2\arrayrulewidth}
Pool	& Level &	Trend 	& Seasonal  & Trend and \\
&  only	&	only	& only & seasonal \\
\Xhline{2\arrayrulewidth}
All models & 2 & 6 & 3 & 8 \\
No multiplicative trend & 2 & 4 & 3 & 6 \\
Damped trend & 2 & 3 & 3 & 4 \\
Match error with seasonal type & 2 & 6 & 2 & 6 \\
Reduced & 2 & 2 & 2 & 2 \\
\Xhline{2\arrayrulewidth}
\end{tabular}
\label{tab:exppoolsmodelssummary}
\end{table}

\subsection{Reduced pools of ARIMA models}\label{sec:pools_arima}

Similar to the reduced pools of exponential smoothing models, we also explored the forecasting performance of subsets of ARIMA models as a function of their complexity. We focused on constraining the maximum order of the candidate ARIMA models so that $p+q \leq K$ for nonseasonal data and $p+q+P+Q \leq K$ for seasonal data, with $K \in \{1, 2 \dots, 8\}$. For each value of $K$, we performed an exhaustive search across all possible models of order $K$ or below and selected the best one based on information criteria (see Subsection \ref{sec:ics}). Table \ref{tab:arimapoolsmodelssummary} provides lists of the ARIMA pools of models considered. For each possible combination of $p$, $q$, $P$, and $Q$, we also fit a model with a constant if $d+D \leq 1$, similar to \citep{Hyndman2008b}. In contrast to the exponential smoothing family, there is always a balance with regard to the auto-regressive (AR) and the moving-average (MA) terms within each ARIMA pool. Also, unlike in the exponential smoothing models, ARIMA pools tend to increase significantly in size as $K$ increases, emphasizing the importance of a balance between performance and computational cost.

\begin{table}[h]\small
\centering
\caption{ARIMA models in each pool.}\vspace{0.25cm}
\begin{tabular}{ccc}
\Xhline{2\arrayrulewidth}
Maximum & No. of  & Examples of models in the pool \\
Order & Models & (in addition to previous pools) \\
\Xhline{2\arrayrulewidth}
$K=1$ & 5 & (0,$d$,0); (1,$d$,0); (0,$d$,1); (0,$d$,0)(1,$D$,0); (0,$d$,0)(0,$D$,1) \\
$K=2$ & 15 & (2,$d$,0); (0,$d$,2); (1,$d$,1); (1,$d$,0)(0,$D$,1); (0,$d$,0)(2,$D$,0)\\
$K=3$ & 35 & (3,$d$,0); (0,$d$,3); (2,$d$,1); (1,$d$,0)(1,$D$,1); (0,$d$,2)(1,$D$,0) \\
$K=4$ & 70 & (4,$d$,0); (2,$d$,2); (1,$d$,3); (2,$d$,1)(0,$D$,1)\\
$K=5$ & 126 & (0,$d$,5); (4,$d$,1); (3,$d$,0)(1,$D$,1); (1,$d$,1)(1,$D$,2) \\
$K=6$ & 210 & (2,$d$,4); (1,$d$,5); (0,$d$,6); (1,$d$,2)(2,$D$,1)\\
$K=7$ & 330 & (5,$d$,2); (0,$d$,7); (1,$d$,2)(3,$D$,1); (3,$d$,3)(0,$D$,1) \\
$K=8$ & 495 & (8,$d$,0); (2,$d$,6); (5,$d$,3); (2,$d$,2)(2,$D$,2)\\
\Xhline{2\arrayrulewidth}
\end{tabular}
\label{tab:arimapoolsmodelssummary}
\end{table}

\subsection{Selecting between models}\label{sec:ics}

The best candidate model is needed without regard for model preferences or the size of the respective pools of candidate models. Selecting the best model for each series is difficult. \cite{Gardner1988-vz} suggested examining the time series' variances of differences as a method to choose among exponential smoothing models. A better approach for selecting among forecasting models is via cross-validation, which follows the process of fixed/rolling origin evaluation \citep{Tashman00}. However, this approach has two significant drawbacks: a series must have enough observations to allow a hold-out sample, and it has high computation costs. Each model will be fitted at least twice: once for validation (even more times for cross-validation) and once for forecasting. On the other hand, cross-validation can be used with both the exponential smoothing and ARIMA families.

Arguably, information criteria are nowadays the most popular approach for model selection in forecasting research. Information criteria select a model based on complexity-penalized maximum likelihood. Maximizing the likelihood is approximately equivalent to maximizing the in-sample fit. Model selection with information criteria is computationally fast and generally robust \citep{Hyndman2002,Hyndman2008-iu,Hyndman2008b}. Popular forecasting packages offer model selection with information criteria by default. The three most widely used approaches are:
\begin{itemize}[noitemsep,nolistsep]
\item Akaike's information criterion, defined as $\text{AIC} = -2\log(L) + 2k$, where $L$ is the likelihood and $k$ is the total number of parameters.
\item Bayesian information criterion, defined as $\text{BIC} = \text{AIC} + k[\log(n) - 2]$, where $n$ is the sample size (number of in-sample observations).
\item Akaike's information criterion corrected for small sample sizes, defined as $\text{AICc} = \text{AIC} + \frac{2k(k+1)}{n-k-1}$. 
\end{itemize}

Note that AIC and AICc have explicit forecasting relevance by being asymptotically optimal for the one-step-ahead out-of-sample mean squared error. This is similar to the concept of one-step-ahead cross-validation. BIC has no such forecasting connection. Furthermore, AICc is the default criterion in the \texttt{ets} function of the \textit{forecast} package of the R statistical software and  closely approximates AIC as the sample size (the number of observations in a series) increases. Thus, the AICc seems the logical choice for model selection \citep{Kolassa2011}. In this study, for each of the various pools of exponential smoothing and ARIMA models presented in Sections \ref{sec:pools_ets} and \ref{sec:pools_arima}, we have selected the "best" model based on AICc values. Using AIC or BIC instead of AICc resulted in very similar insights, which confirms the findings by \cite{Billah2006}.

\subsection{Measuring performance and computational cost}\label{sec:performance}

We measured both performance and computational cost for each of the pools of models. For performance, we considered both point forecast accuracy and prediction interval coverage. Point forecast accuracy is measured using the Mean Absolute Scaled Error (MASE). MASE is equal to the mean absolute error scaled by the in-sample one-step-ahead mean absolute error of seasonal Naive and has better statistical properties than the widely-used percentage error measures \citep{Hyndman06,Franses2016-pj}. MASE is defined as
$$
\text{MASE} = \frac{1}{h} \frac{ \sum_{t=n+1}^{n+h} {|y_{t}-f_{t}|} } {\frac{1}{n-s} \sum_{i=s+1}^{n} |y_{i}-y_{i-s}|},
$$
\noindent where $y_t$ is the actual observation at time period $t$, $f_t$ is the respective forecast, $h$ is the forecasting horizon, and $s$ is the number of periods within a seasonal cycle (e.g., $s=12$ for monthly data). The values of MASE are averaged across different time series. Although forecasters often use percentage error measures, we decided not to report them because of their asymmetry \citep{Goodwin1999} and because the insights gained from them generally support our MASE results.

The performance of the prediction intervals is measured by the Mean Scaled Interval Score \citep[MSIS,][]{Makridakis2019-oy}. The interval score \citep{Gneiting2007-oj} considers both the interval width and instances in which actual values fall outside the interval range. It is scaled similarly to the MASE to allow averaging across series. The MSIS is defined as
$$
\text{MSIS} = \frac{ W(l_t, u_t, y_t) } {\frac{1}{n-s} \sum_{i=s+1}^{n} |y_{i}-y_{i-s}|},
$$
\noindent where $l_t$ and $u_t$ are the lower and upper prediction intervals for period $t$ and
$$
W(l_t, u_t, y_t) = \begin{cases} (u_t - l_t) & l_t \leq y_t \leq u_t \\ (u_t - l_t) + \frac{2}{\alpha}(l_t - y_t) & y_t < l_t \\ (u_t - l_t) + \frac{2}{\alpha}(y_t - u_t) & y_t > u_t, \end{cases}
$$
\noindent where $(1-\alpha) \times 100\%$ is the confidence level. 

Computational cost is the time (in seconds) needed to fit all relevant models and estimate the prediction intervals of the best one, averaged across time series. We used a machine equipped with a Microsoft Windows 11 Pro operating system, an Intel Core i7-8700 processor clocked at 3.20GHz, providing six cores and 16GB of memory. To compute the forecasts for this study, we applied parallelization, with each series being processed on a single core. 

All calculations for this study were performed using the R statistical software and the \textit{forecast} package. Although other platforms (C++ or Python) could provide faster implementations, we do not think that the relative differences in computational cost for the different pools of models (Sections \ref{sec:pools_ets} and \ref{sec:pools_arima}) would change significantly.

\subsection{A machine learning benchmark}\label{sec:lightgbm}

To benchmark our results more objectively, we considered a state-of-the-art machine learning algorithm, namely LightGBM, that performs nonlinear regression using gradient-boosted trees \citep{NIPS2017_6907}. This method is considered the method of choice in Kaggle's recent forecasting competitions. Many winners of these competitions, including the retail-focused "M5", "Corporación Favorita Grocery Sales Forecasting", and "Recruit Restaurant Visitor Forecasting", have used this method \citep{BOJER2020}. In brief, LightGBM generates multiple independent regression trees in sequence to decrease the error made by the previously trained tree \citep{FREUND1997119}. Apart from having a bigger learning capacity than single trees, LightGBM is more robust to noise and less likely to overfit on the training data \citep{FRIEDMAN2002367} while also balancing accuracy and computational speed through a random sampling technique that selects the optimal split in each node.

A single, global  forecasting model \citep{JANUSCHOWSKI2020167} was trained across the complete data set and used to simultaneously predict the respective series. As for machine learning models, cross-learning approaches are generally preferred over local ones, trained in a series-by-series fashion, because they typically result in similar - if not better - accuracy, lower computational cost, and fewer issues related to data availability \citep{SEMENOGLOU20211072}. The standard method of constant size with rolling input and output windows \citep{SpiliotisORIJ20372} was used to train the model. In this method, the in-sample data of each series is split into multiple input and output vectors of sizes $w_{in}$ and $w_{out}$, respectively, with each output vector containing the observations that succeed those of the corresponding input vector. Thus, for each series of length $n$, a total of $n-w_{in}-w_{out}+1$ train samples can be created. After training the model with these samples, the last $w_{in}$ observations of the series can be used as input to forecast the following $w_{out}$ periods. To simplify the benchmark design, and because some of the series are seasonal, $w_{in}$ was set equal to $2 \times h$ to ensure the look-back windows were created long enough to capture possible trends and seasonal patterns. Conversely, given that standard decision-tree-based models provide a single output, $w_{out}$ was set equal to one, a setting that also facilitates training for short series of limited historical observations. According to this setup, multiple-step-ahead forecasts can be computed recursively \citep{BENTAIEB20127067}. To facilitate training across multiple series of different scales, a min–max transformation was applied to each of the samples created to render their scale in a target range of [0, 1]. In total, the described method resulted in about 8.8 million training samples. 

Given that LightGBM uses several hyper-parameters\footnote{https://lightgbm.readthedocs.io/en/latest/index.html} that can affect forecasts, we determined their values using grid search, an automated method that explores a set of different hyper-parameter values and computes forecasting performance on a predefined validation set to find the optimal ones by minimizing an accuracy measure of preference. (In this case, the preferred measure was the mean absolute error because it better matches the statistical properties of MASE used to evaluate post-sample performance.) We focused on the most critical hyper-parameters of LightGBM, selecting the shrinkage rate (\textit{learning\_rate}) from [0.01, 0.025, 0.05, 0.1], the number of boosting iterations (\textit{num\_iterations}) from [100, 500, 1000, 5000], the maximum number of leaves in one tree (\textit{num\_leaves}) from [64, 256, 1024], and the minimum number of data in one leaf (\textit{min\_data\_in\_leaf}) from [8, 16, 32, 64, 128]. For the sake of simplicity, the validation set consisted of the last observation included in the series in the training set. The selected hyper-parameters were \textit{learning\_rate}=0.05, \textit{num\_iterations}=5000, \textit{num\_leaves}=1024, and \textit{min\_data\_in\_leaf}=128.

Similar to most machine learning models, LightGBM provides only point forecasts. The method does not estimate probabilistic forecasts (e.g. quantile or interval forecasts) in an automated, widely accepted way and requires additional empirical computations or simulations to do so. We will therefore limit the comparisons made with this benchmark to point forecasts.

\subsection{Data}\label{sec:data}

We use the monthly data from the M, M3 and M4 forecasting competitions \citep{Makridakis82,Makridakis2000a,Makridakis2019-oy}. In total, we considered an extensive sample of 50,045 real-life series. The series exhibits various lengths, but the forecasting horizon is fixed at 18-periods-ahead. Following the design of the M competitions, we produced forecasts once for each model and series combination, using a fixed-origin evaluation and withholding the appropriate number of observations. We also used the daily retail data from the M5 forecasting competition \citep{MAKRIDAKIS2021m5data} as a case study in Section \ref{sec:case_study}.

\section{Empirical results}\label{sec:evaluation}

\subsection{Forecasting performance and cost}\label{sec:performance_cost}

Tables \ref{tab:resultsmainexpsmoothing} and \ref{tab:resultsmainArima}, respectively, present the performance and cost comparisons for the exponential smoothing and ARIMA pools considered. In each table, the  rows present the performance of the pools of models described in Section \ref{sec:design}. The columns refer to the three performance measurements, MASE, MSIS, and computational cost (seconds per series). For the MSIS, we considered a 95\% confidence level ($\alpha = 0.05$). The best result for each measure and data frequency is in boldface.

\begin{table}[h]\small
\centering
\caption{Average forecasting performance and computational cost results of five exponential smoothing pools.}\vspace{0.25cm}
\begin{tabular}{cccccc}
\Xhline{2\arrayrulewidth}
Pool	&	MASE	& MSIS & Cost \\
\Xhline{2\arrayrulewidth}
All models& 1.046 & 2374.937  & 1.268\\
No multiplicative trend & 0.947  & 8.258  & 0.797 \\
Damped trend & 0.979 & 14.922  & 0.898 \\
Match error with seasonal type & 1.047 & 2374.942 & 1.023 \\
Reduced & \textbf{0.942} & \textbf{8.201} & \textbf{0.365} \\
\Xhline{2\arrayrulewidth}
\end{tabular}
\label{tab:resultsmainexpsmoothing}
\end{table}

\begin{table}[h]\small
\centering
\caption{Average forecasting performance and computational cost for different maximum order of ARIMA models.}\vspace{0.25cm}
\begin{tabular}{cccccc}
\Xhline{2\arrayrulewidth}
Maximum Order	&	MASE	& MSIS & Cost \\
\Xhline{2\arrayrulewidth}
$K=1$  & 0.962 & 8.969 & \textbf{0.027} \\
$K=2$  & 0.938 & \textbf{8.727} & 0.144 \\
$K=3$  & 0.933 & 8.745 & 0.850 \\
$K=4$  & \textbf{0.931} & 8.762 & 4.281 \\
$K=5$  & 0.932 & 8.800 & 16.065 \\
$K=6$  & 0.935 & 8.891 & 48.102 \\
$K=7$  & 0.938 & 8.970 & 119.458 \\
$K=8$  & 0.941 & 9.111 & 255.347 \\
\Xhline{2\arrayrulewidth}
\end{tabular}
\label{tab:resultsmainArima}
\end{table}

Comparing performance across the different exponential smoothing pools (Table \ref{tab:resultsmainexpsmoothing}) suggests ``all models'' result in the worst performance and the highest computational times. Recall that lower scores are better, both for performance and cost. In some cases, the forecasts and prediction intervals produced are explosive, leading to very poor predictions and substantial forecasting errors, resulting in very high values of the arithmetic means, especially those for the MSIS (across all frequencies). This poor performance of ``all models'' suggests that identifying the best model from a large pool of candidates is sometimes influenced by overfitting and model selection that tend to produce unrealistic out-of-sample forecasts.

The major problem of the ``all models'' pool is that it allows multiplicative trend models. If these are excluded (``no multiplicative trend'' pool), then performance across all measures is significantly better. In fact, the second pool of models results in the second-best performance on average. More importantly, excluding the four multiplicative trend models leads to a 37\%  reduction in computational time compared to the time needed for the ``all models'' pool.

The next pool of models considered allows only models with a damped trend when a trend component is included. This pool results in superior performance compared with ``all models'' and needs less time to produce forecasts for seasonal frequencies. Although some multiplicative trended models are still allowed, the outlying cases are moderated by forcing a damped trend. 

Matching the error type with the type of seasonality essentially blocks multiplicative errors with additive seasonality models from consideration (because additive error and multiplicative seasonality were already excluded). This elimination of three models has no effect on performance, which is on par with that of the ``all models'' pool, but it slightly trims computational time.

Finally, we evaluated the ``reduced'' pool of exponential models. This is the smallest of the exponential smoothing pools considered, consisting of only eight models, and is the only one with a balanced selection of exponential smoothing in terms of their components (trend and/or seasonality). The reduced pool results in the best performance overall. Furthermore, this pool offers computational times 54\% lower than the second-fastest ``no multiplicative trend'' pool (which already offered significant computational gains).

Next, we consider the ARIMA results, in which the various pools of models are defined with regard to the maximum order allowed. As discussed in \ref{sec:pools_arima}, we considered pools of models with maximum orders between 1 and 8. For point forecast accuracy, the results follow a U-shaped function, and the best accuracy is achieved with $K=4$. Increasing the maximum order does not improve average accuracy, but decreasing the maximum order to $K=3$ or $K=2$ lowers accuracy slightly. The uncertainty results (MSIS) show best performance is achieved for $K=2$, with the uncertainty estimation progressively decreasing as $K$ increases. Large models do not gain forecasting performance, possibly because information criteria cannot properly penalize for overfitting.

Overall, better performance in the ARIMA family of models is not necessarily correlated with an increase in the maximum order of the candidate models. In fact, allowing for large models leads to poor average performance. Moreover, the computational cost is also considerably increased: A maximum order of $K=8$ is 300 times slower compared with $K=3$.

Comparing the results in Tables \ref{tab:resultsmainexpsmoothing} and \ref{tab:resultsmainArima} shows ARIMA's forecasts are more accurate. ARIMA, however, performs worse in estimating uncertainty. The best-performing exponential smoothing pools significantly outperform ARIMA pools, achieving better MSIS scores. Lastly, the reduced exponential smoothing pool is computationally faster than the ARIMA pools for $K>2$.

We also identified the drivers of higher computational cost within each of the two families of models. The need to perform simulations to derive prediction intervals (because some models lack analytical expressions) is the most computationally demanding factor for exponential smoothing models, followed by the inclusion of seasonality and a trend component. In an ARIMA model, the main culprit for computational cost is higher seasonal autoregressive or moving average orders ($P$ and $Q$), followed by nonseasonal orders ($p$ and $q$). The length of a series had a lesser effect in both model families.

To test the statistical significance of the differences between pools, we used the modified Diebold-Mariano (DM) test to compare the performance of the various pools of models to rule out that the lower scores were due to chance \citep{Harvey1997-xv, Athanasopoulos2009-db}. The DM test allows comparisons of forecasting models and accounts for the autocorrelation that exists when forecasting $h$-steps ahead of the origin. A DM test permits calculating the percentage of times a pool of models is significantly better or worse than other pools for each forecast horizon.

With exponential smoothing models, when accuracy measured by MASE is considered, the reduced pool is never worse than other pools and is significantly better 50\% to 100\% of the time. In considering the uncertainty performance as measured by MSIS, the reduced pool is significantly better than other pools from 28\% up to 66.7\% of the time and never significantly worse.

For the ARIMA models, the lower maximum order models ($K=1$) are statistically significantly worse than higher maximum order models, especially on MASE. Models of a maximum order of $K=4$ and 2 are, however, never significantly worse than other order models while also significantly better than models with $K \geq 6$ in 61\% to 94\% of the time for accuracy (MASE) and uncertainty (MSIS), respectively.

\subsection{Component analysis}\label{sec:forecast_profiles_analysis}

In this subsection, we explore which model is most frequently selected across each pool of models. Specifically, we compare the frequency with which the various components are selected. For the exponential smoothing family, we first calculated the relative percentage frequencies for which each of the four basic forecast profiles (level only, trend only, seasonal only, and trend and seasonal) is selected as optimal. We compared the relative frequencies that different model categories are selected when using ``all models'' versus the ``reduced'' pools. Table \ref{tab:featuresselectedETS} presents the results. 

We noticed that the relative frequencies of trended-only or trended and seasonal models drop significantly when we move from the ``all models'' pool to the ``reduced'' one. At the same time, the ``reduced'' framework selects more often level-only models and slightly more often selects seasonal (but not trended) models. A more in-depth analysis reveals that for about 20.5\%  of the series in which a trended model was suggested by an ``all models'' pool, the ``reduced'' pool suggests a level model as optimal. Similarly, 7.3\% of the series best fitted with trended and seasonal models under ``all models'' became fitted by an only-seasonal model. Overall, the ``reduced'' pool results in not only better  performance but also the selection of simpler models.

\begin{table}[h]\small
\centering
\caption{The relative percentage frequencies with which exponential smoothing models with different components are selected.}\vspace{0.25cm}
\begin{tabular}{ccc}
\Xhline{2\arrayrulewidth}
Model	components&			All models	& Reduced\\	
\Xhline{2\arrayrulewidth}
Level		                &  25.1 & 30.9  \\
Trended	               	&  25.8 & 20.3  \\
Seasonality	            	&  17.0 & 18.1  \\
Trend \& Seasonality		&  32.2 &  30.6  \\
\Xhline{2\arrayrulewidth}
\end{tabular}
\label{tab:featuresselectedETS}
\end{table}

The respective results for the ARIMA pools of models are reported in Table \ref{tab:featuresselectedARIMA}. We present the percentage of times that the selected models in the pool with maximum order $K$ differ from the previous pool ($K-1$) in terms of the order of $p$, $q$, $P$, $Q$, or any of the above. The percentage of changes for a model reduces from 77.4\% for $K=2$ to 32.3\% for $K=8$. This suggests the possibility of finding a better model decreases as $K$ (and computational time) increases.

\begin{table}[h]\small
\centering
\caption{The relative percentage frequencies of the differences in the order of the selected ARIMA models compared with the previous ($K-1$) pool of models.}\vspace{0.25cm}
\begin{tabular}{cccccccc}
\Xhline{2\arrayrulewidth}
Orders &			\multicolumn{7}{c}{Maximum order ($K$)}\\	
 & 2 & 3 & 4 & 5 & 6 & 7 & 8 \\
\Xhline{2\arrayrulewidth}
$p$	& 19.4 & 20.9 & 20.6 & 20.0 & 18.4 & 16.8 &  15.5 \\
$q$	& 27.2 & 26.2 & 23.7 & 22.3 & 20.2 & 18.3 &  16.6 \\
$P$	& 20.4 & 20.9 & 19.5 & 17.1 & 15.6 & 13.9 &  13.0 \\
$Q$	& 23.6 & 20.8 & 20.2 & 18.1 & 17.3 & 15.6 &  13.9 \\
Any   & 77.4 & 64.7 & 54.7 & 46.9 & 41.2 & 36.4 &  32.3 \\
\Xhline{2\arrayrulewidth}
\end{tabular}
\label{tab:featuresselectedARIMA}
\end{table}

\subsection{Uncertainty estimation analysis}\label{sec:uncertainty_estimation_analysis}

\begin{figure}[!ht]
    \centering
    \includegraphics[width=5.5in]{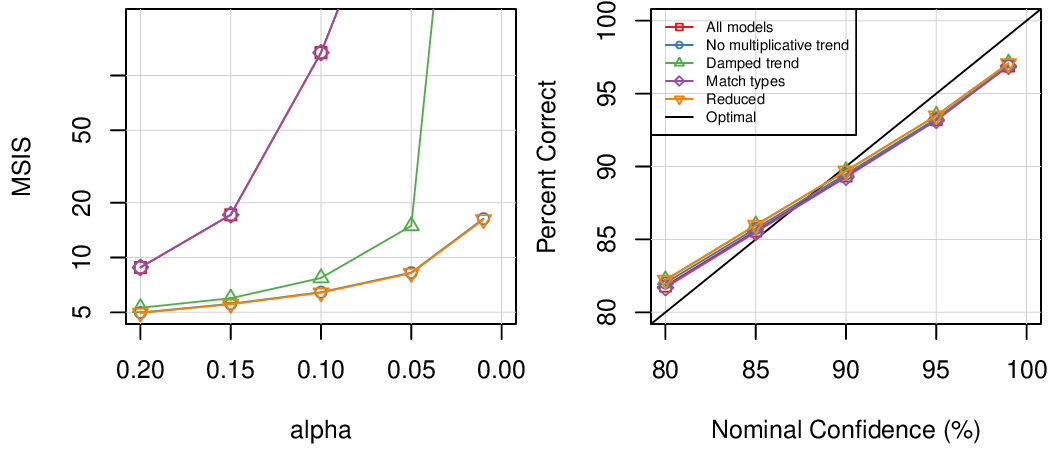}
    \caption{The MSIS performance and calibration for different confidence levels and the exponential smoothing pools of models. The MSIS for the ``all models'' pool (red line) is not depicted because of the very large values. The vertical axes of the MSIS plots are log-scaled.}
    \label{fig:MSISCalCovETS}
\end{figure}

\begin{figure}[!ht]
    \centering
    \includegraphics[width=5.5in]{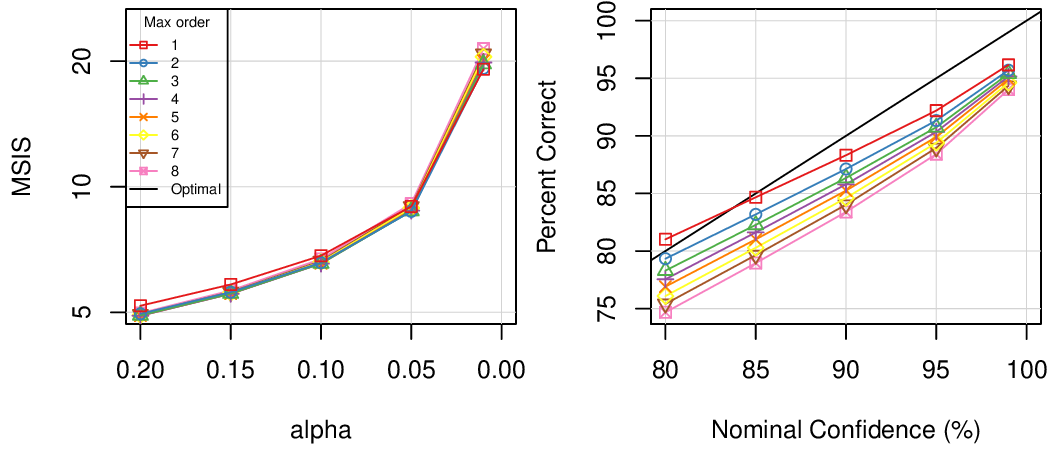}
    \caption{The MSIS performance and calibration for different confidence levels and the ARIMA pools of models. The vertical axes of the MSIS plots are log-scaled.}
    \label{fig:MSISCalCovARIMA}
\end{figure}

The MSIS results in Tables \ref{tab:resultsmainexpsmoothing} and \ref{tab:resultsmainArima} assumed a 95\% confidence level ($\alpha = 0.05$). We now expand the analysis to more quantiles: 80\%, 85\%, 90\%, 95\%, and 99\%. These quantiles refer to the nominal confidence expected from the prediction intervals we obtained. Figures \ref{fig:MSISCalCovETS} and \ref{fig:MSISCalCovARIMA} show the MSIS for exponential smoothing and ARIMA, respectively, for more confidence levels (80\%, 85\%, 90\%, 95\%, and 99\%), for each pool of models (left panels of each figure). We also report the calibration (right panels), which shows the percentage of times that the actual observations are not outside the lower and upper prediction intervals. 

With regard to the exponential smoothing results (Figure \ref{fig:MSISCalCovETS}), the MSIS values for the ``reduced'' pool of models are lower or equal to that of any other pool of models considered (they are virtually identical to the ``no multiplicative trend'' pool). All pools of models have similar miscalibration issues. In any case, the ``reduced'' pool of models always offers the best calibration of all pools. Overall, we can see that the good performance of the ``reduced'' pool in estimating the uncertainty around the forecasts holds for various confidence levels.

The ARIMA results (Figure \ref{fig:MSISCalCovARIMA}) indicate that a maximum order of two or three offers the best MSIS results, regardless of the value of the confidence level. Moreover, a maximum order equal to one results in the best calibration, even if all pools of models are significantly undershooting. We saw in Table \ref{tab:resultsmainArima} that lower maximum orders are outperformed in the monthly data regarding uncertainty estimation based on MSIS. Although this is true for all confidence levels, small pools of ARIMA models are better than larger pools in terms of calibration; this difference is amplified as we consider lower nominal confidence levels.

\subsection{Other pools of exponential smoothing models}\label{sec:other_pools_models}

Although the various pools of ARIMA models are defined naturally based on their maximum order, this is not true for the five exponential smoothing pools considered in this study. It can be argued that our definition of the reduced pool is arbitrary. In this subsection, we now expand our analysis to a much larger number of exponential smoothing pools. The only constraint we enforce is that each pool has at least one model for each of the four basic forecast profiles, as presented in Table \ref{tab:exppoolsmodelssummary}: level only, trend only, seasonal only, and trend and seasonal.

\begin{figure}[!ht]
    \centering
    \includegraphics[width=4.5in]{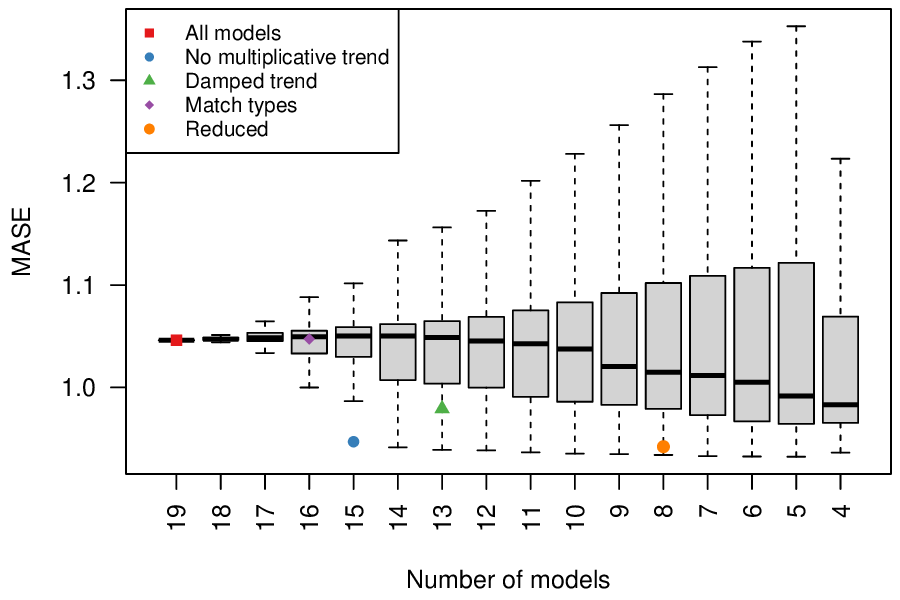}
    \caption{The performance of each possible pool of exponential smoothing models.}
    \label{fig:FullRun}
\end{figure}

We analyze the performance of 337,365 seasonal pools (each containing at least four models with a maximum of 19 models). For each pool size (4 to 19), we depict in Figure \ref{fig:FullRun} the box plot of the average MASE performances of the respective pools of models. We also present the performance of the five base pools defined in Section \ref{sec:pools_ets}. For example, the reduced pool of models (orange dot) performs very competitively against others of equal size (at the lower end of the respective box plots, reflecting a lower MASE) but also robustly across pools of all sizes. In fact, the reduced pool of models ranks in the top 1\% across all possible pools. Similar insights are obtained for MSIS as a performance metric.

\subsection{Forecast-value-added analysis}\label{sec:FVA_analysis}

\cite{Gilliland2013-dw,GillilandReport2015} suggested that the forecast-value-added (FVA) of each step in the forecasting process should be measured when progressing from simple to more complex procedures. Simply put, FVA is the percentage improvement in forecasting performance achieved by each added step. FVA can be seen as an underfitting versus overfitting exercise in which one attempts to identify the optimal level of complexity in a forecasting process to achieve the desired levels in performance. If more complex processes add forecasting value, then simpler approaches might be under-fitted and require improvement. Conversely, if more complex approaches offer no performance gains, this might indicate overfitting and that the simpler approaches might be good enough.

We conducted a FVA analysis for both the exponential smoothing and ARIMA families. With the exponential smoothing family, we used three of the five pools of models previously considered: reduced (eight models), no multiplicative trend (15 models), and all models (19). In ARIMA, we considered the pools where the maximum order is lower or equal to 5. We sorted the pools of models within each family based on their size in an attempt to map the gains at each step of increasing complexity. At the same time, we expanded the FVA analysis to include what we refer to as the computational cost reduction (CCR) of each step. Our analysis is presented in Tables \ref{tab:fva-monthly-ets} and \ref{tab:fva-monthly-arima} in which forecasting performance is measured by MASE. Positive values in FVA suggest more sophisticated approaches offer performance gains over simpler ones. Positive values in CCR indicate a decrease in the computational cost. 

\begin{table}[h]\footnotesize
\centering
\caption{Forecast-value-added (FVA) and computational-cost-reduced analysis for the monthly data frequency and the exponential smoothing family of models.}\vspace{0.25cm}
\begin{tabular}{lcccccc}
\Xhline{2\arrayrulewidth}
Pool of models & MASE & FVA & FVA & Cost per series & CCR & CCR \\
(sorted by complexity) &  & vs. 1 & vs. 2 & (seconds) & vs. 1 & vs. 2 \\
\Xhline{2\arrayrulewidth}
1: Reduced & 0.942 &  &  & 0.365 &  &  \\
2: No multiplicative trend & 0.947 & $-$0.5\% &  & 0.797 & $-$118\% &  \\
3: All models & 1.046 & $-$11\% & $-$10.5\% & 1.268 & $-$247\% & $-$59\% \\
\Xhline{2\arrayrulewidth}
\end{tabular}
\label{tab:fva-monthly-ets}
\end{table}

\begin{table}[h]\footnotesize
\centering
\caption{Forecast-value-added and computational-cost-reduced analysis for the monthly data frequency and the ARIMA family of models.}\vspace{0.25cm}
\begin{tabular}{ccccccccccccc}
\Xhline{2\arrayrulewidth}
Maximum & MASE & FVA & FVA & FVA & FVA & Cost per series & CCR & CCR & CCR & CCR \\
Order &  & vs. 1 & vs. 2 & vs. 3 & vs. 4 & (seconds) & vs. 1 & vs. 2 & vs. 3 & vs. 4 \\
\Xhline{2\arrayrulewidth}
1: $K=1$ & 0.962 & & &  &  &  0.027 & & &  &  \\
2: $K=2$ & 0.938 & 2.5\% & &  &  &  0.144 & $-$433\% & &  &  \\
3: $K=3$ & 0.933 & 3.1\% & 0.5\% &  &  &  0.850 & $-$3048\% & $-$490\% &  &  \\
4: $K=4$ & 0.931 & 3.2\% & 0.7\% & 0.1\% &  &  4.281 & $-$15756\% & $-$2873\% & $-$404\% &  \\
5: $K=5$ & 0.932 & 3.1\% & 0.6\% & 0.1\% & 0\% &  16.065 & $-$59400\% & $-$11056\% & $-$1790\% & $-$275\% \\
\Xhline{2\arrayrulewidth}
\end{tabular}
\label{tab:fva-monthly-arima}
\end{table}

Table \ref{tab:fva-monthly-ets} shows that expanding the collection of exponential smoothing models not only decreases the FVA but also substantially increases the CCR. Table \ref{tab:fva-monthly-arima} shows that although the CCR is always negative (suggesting increased cost as more models are considered), some forecast value is added; this is true up until $K=4$. Although $K=4$ performs slightly more accurately than $K=3$, this improvement comes with significant added computation (more than four times more expensive). We suggest that the FVA should be balanced against the CCR to make an informed decision on the value of the maximum order.

\subsection{Comparisons with the machine learning benchmark}\label{sec:FVA_analysis}

We now focus on the performance of the reduced forms of the two time series forecasting families of models examined, both in terms of forecasting accuracy and computational cost, compared with those of the machine learning benchmark introduced in Subsection \ref{sec:lightgbm}. This includes the reduced form of ETS and the implementation of the ARIMA approach for $K=4$. The results are summarized in Table \ref{tab:resultscompml}. We should note that the LightGBM approach is unable to produce quantile forecasts. As such, the comparison below explicitly focuses on point forecast accuracy (MASE). Also, note that the computational cost of LightGBM includes all steps required to produce the final forecasts, i.e., feature engineering (creation of look-back windows), hyper-parameter optimization, training, and inference.

\begin{table}[h]\small
\centering
\caption{Average forecast accuracy and computational cost of the reduced forms of two time series forecasting families of models (ETS and ARIMA) compared with those of the machine learning benchmark (LightGBM).}\vspace{0.25cm}
\begin{tabular}{cccccc}
\Xhline{2\arrayrulewidth}
Approach	&	MASE &	Cost \\
\Xhline{2\arrayrulewidth}
ETS - Reduced & 0.942 & \textbf{0.365} \\
ARIMA - $K=4$ & 0.931 & 4.281 \\
LightGBM & \textbf{0.921} & 7.724 \\
\Xhline{2\arrayrulewidth}
\end{tabular}
\label{tab:resultscompml}
\end{table}

We observed that forecasting performance between the three approaches, reduced ETS, ARIMA with $K=4$, and LightGBM, is close. Still, LightGBM outperforms ETS and ARIMA in terms of MASE by margins of 1.2\% and 2.2\%, respectively. However, LightGBM is significantly slower than either ARIMA or ETS, with reduced ETS more than 20 times faster than LightGBM, while ARIMA is 45\% faster than LightGBM. Overall, there is a clear trade-off between forecasting accuracy and computational cost that needs to be considered. In the next section, we investigate this trade-off, focusing on a retail data set, a context in which the computational burden is of major importance, given the large number of forecasted items.


\section{Case study: Walmart}\label{sec:case_study}

Our proposed reduced pools of models are particularly important when numerous forecasts must be prepared frequently. Retailers face this challenge because they must produce forecasts for hundreds of thousands of SKUs for each of their locations/stores. The total number of forecasts needed at the most granular levels can be in the hundreds of millions, if not billions of forecasts at every review cycle. Brian Seaman, Senior Director of Data Science at Walmart Labs, noted ``the computational burden of heavily multi-model approaches may become too great'' \citep{Seaman2021}. \cite{Yelland2019} discussed the challenge of forecasting at scale for Target and noted that Target should be willing to take a small hit in forecasting performance in order to improve computational cost and the ability to explain models. Finally, Stephan Kolassa, a Data Science Expert at SAP Switzerland AG, writes: ``If you are tempted to invest heavily in data scientists and expect them to work wonders, make sure to compare their methods to simple benchmarks that you can probably implement at a fraction of the cost of an ML pipeline'' \citep{Kolassa2021M5}.

We will apply our insights from the previous empirical section to a subset of Walmart's data representing three product categories (hobbies, food, and household), divided into seven departments, and containing sales for 3,049 SKUs. These data were recently used in the M5 forecasting competition \citep{Makridakis2020-wq,Makridakis2020-bj}. The SKU sales are available for ten stores in three U.S. states (California, Texas, and Wisconsin), so we can access 30,490 time series at the SKU-Store level. The data are daily and cover about 5.5 years (1,969 days), but some series are significantly shorter \citep[for more details on the M5 data set, see][Section 4]{MAKRIDAKIS2021m5data}

The data are organized in a grouped hierarchy with levels consisting of sales for the total, each state, store, category, department, and product. For this case study, we focused on and evaluated forecasts at the most granular level (SKU-Store). This level is most relevant for inventory control, which depends on accurate forecasts to reduce stockout costs, inventory holding costs, and wastage \citep{Seaman2021}.

Similar to Section \ref{sec:evaluation}, we benchmarked our results by using a special implementation of LightGBM, to be called ``LightGBM ensemble,'' that closely replicates the winning method of the M5 ``Accuracy'' challenge \citep{IN2021}. Variants of this approach were widely considered by the teams that participated in M5, including most of the top-performing ones. The winning submission outperformed univariate exponential smoothing, the top-performing benchmark of M5, by an impressive 22.4\%, according to the official measure of the competition that summarized forecasting accuracy across 12 cross-sectional levels. This improvement, however, dropped to 3.4\% at the most granular level at which the data is noisy and intermittent and the forecasting task is at its most challenging. In this regard, it is within the scope of the present section to explore the potential value of exponential smoothing and its reduced forms over this state-of-the-art, top-performing method, both in terms of accuracy and computational cost. Our analysis suggests that reduced pools of simple univariate forecasting methods are capable of accelerating forecasting without significant losses in accuracy.

Specifically, the LightGBM ensemble considers various decision-tree-based models trained by using data pooled per store (10 models), store-category (10 stores $\times$ 3 categories = 30 models), and store-department (10 stores $\times$ 7 departments = 70 models). The forecasts of these models, each specialized in forecasting specific SKU-Store series, are then combined appropriately using equal weights. For instance, in order to forecast the sales of a food product sold in a California store, forecasts of three models are combined, the first trained across all products sold in that store, the second trained across all the food products sold in that store, and the third trained across all the food products of the same department sold in that store. Note that the winner of the ``Accuracy'' track of the M5 forecasting competition originally considered two strategies for generating forecasts from each model, the first being of recursive nature and the second of non-recursive nature \citep{Bontempi2013}. We do not consider recursive models in more detail for the sake of simplicity. Using such models did not improve accuracy significantly in preliminary tests, but it did increase computational cost. As for the features used as input, the models considered identifiers about the product, store, product-category, product-department being forecasted, calendar-related information (e.g., day of the week, week of the year, and month), special days and holidays, and past sales.

We compared the performance of the LightGBM ensemble with the exponential smoothing family of models that corresponds to the best benchmark used at the ``Accuracy'' track of the M5 forecasting competition. In fact, univariate exponential smoothing forecasts produced at the most granular level and consequently aggregated (bottom-up hierarchical approach) were able to outperform 92.5\% of the 5507 teams that participated in the M5 ``Accuracy'' challenge \citep{Kolassa2021M5}, an impressive result for an automated univariate forecasting method. 


For our results to be as realistic as possible, accounting for calendar effects and data uncertainty, we used a rolling-origin evaluation process that assesses forecasting performance across a complete year \citep{Tashman00}. Specifically, from the 30,490 series available in the M5 data set, we retained those that contained more than two years of data (28,298 series), that is, a minimum of a year for training and a complete year for testing. Initially, we used the first 1,605 days of data (or less for shorter series) to train the models for forecasting the following 28 periods. Then, we moved the forecasting origin forward by 28 periods and used 1,633 days of data to re-train the models and forecast the next 28 days. This process was repeated 13 times, in other words, until no more data were available for testing purposes. After completing the simulation, we summarized the forecasting accuracy of each forecasting approach by averaging the forecast errors reported for each series and evaluation round. We also tracked computational times and monitored the average time (seconds) elapsed to forecast each series.

Forecasting accuracy was measured using the Root Mean Squared Scaled Error (RMSSE), defined as follows:
\begin{align*}
\text{RMSSE} = \sqrt{ \frac{ \frac{1}{h} \displaystyle\sum_{t=n+1}^{n+h} ({y_{t}-\hat{y}_{t}})^2 } {\frac{1}{n-1} \displaystyle\sum_{t=2}^{n} (y_t-y_{t-1})^2} }.
\end{align*}

Like MASE, RMSSE is independent of the scale of the data, has predictable behavior, has a defined mean and a finite variance, and is symmetric in the sense that it penalizes equally positive and negative forecast errors as well as errors of all sizes. The choice for this specific measure over MASE is justified by the nature of the SKU-Store data: The series are intermittent, involving sporadic unit sales with many zeros, meaning that squared errors are more appropriate for identifying methods that accurately predict the mean of the demand \citep{KOLASSA2016788} instead of the median \citep{Neil1990asfda}. Also, this measure is consistent with the one used in the M5 competition to evaluate the accuracy of the submitted methods. 

Similar to Section \ref{sec:evaluation}, we used the \texttt{ets()} function of the R \textit{forecast} package to produce forecasts for the complete and reduced pools of exponential smoothing models, as discussed in Subsection \ref{sec:pools_ets}, assuming a daily frequency ($s=7$). Because multiplicative models are not defined for series that involve zero values, this limits the search for an optimal exponential smoothing model to the six additive ones (ANN, ANA, AAN, AAA, AAdN, and AAdA) for the ``all models'', ``no multiplicative trend'', and ``match error with seasonal type'' pools and to only four additive models (ANN, ANA, AAdN, and AAdA) for the ``damped trend'' and ``reduced'' pools. Therefore, we will report results only for the ``all models'' and ``reduced'' pools of models. Table \ref{tab:resultswalmart} reports the results from applying these two pools of models from the exponential smoothing family. 

\begin{table}[h]\small
\centering
\caption{Average forecasting accuracy and computational cost results of the exponential smoothing pools of models and the LightGBM ensemble for the Walmart case study.}\vspace{0.25cm}
\begin{tabular}{ccc}
\Xhline{2\arrayrulewidth}
Method	&	RMSSE	& Cost per series\\
& & (seconds)\\
\Xhline{2\arrayrulewidth}
LightGBM Ensemble & 0.717 & 0.710 \\
ETS - All models & \textbf{0.713} & 0.603 \\
ETS - Reduced & 0.714 & \textbf{0.416} \\
\Xhline{2\arrayrulewidth}
\end{tabular}
\label{tab:resultswalmart}
\end{table}

Both exponential smoothing and its reduced pool of models result in slightly more accurate forecasts than the LightGBM ensemble, improving accuracy by 0.57\% and 0.42\%, respectively. Moreover, although the reduced pool of models is slightly less accurate than the complete one, its computational cost is 31\% less. As noted earlier, the search space of the former pool consists of 33\% fewer models than the latter pool. More importantly, exponential smoothing and reduced exponential smoothing decrease computational cost compared to the LightGBM ensemble by an impressive 15\% and 41\%, respectively. We conclude that when we apply reduced pools of exponential smoothing models to large sets of retailers' data, we can maintain (if not improve) the quality of forecasts while significantly reducing the cost of generating them. 

Note that the LightGBM method submitted by the winner of the ``Accuracy'' challenge in the M5 competition was slightly more accurate (by about 3\%) than the exponential smoothing benchmark considered by the organizers. Our results in Table \ref{tab:resultswalmart} are different. Two likely explanations for this difference are that (1) the RMSSE measure used in evaluating accuracy in the M5 competition weighs series based on past sales, whereas our measure weighs series equally, and (2) the M5 results are based on a single evaluation window of data, but our results are based on 13 non-overlapping windows.

We can translate such computational gains into possible monetary savings by considering the adoption of the proposed reduced exponential smoothing pool in large retailers and making some realistic assumptions. \cite{Seaman2017} mentioned that the 5,000 physical stores of Walmart can hold over 200 thousand SKUs each, requiring a total of 1 billion SKU-Store forecasts and 197 thousand CPU-hours\footnote{Note that CPU-hour is not a direct measure of time because parallelization can be applied.} if the LightGBM ensemble is used or 168 thousand CPU hours if all exponential smoothing models were estimated. This computational time would decrease to about 116 thousand CPU hours if the reduced framework of exponential smoothing were used instead.

If Walmart used cloud computing to produce its forecasts, then a typical CPU-hour would cost about \$0.05 \citep{Nikolopoulos2018-wa}. A single cycle of forecast production would cost \$9,850 for computing LightGBM, \$8,400 to estimate all possible exponential smoothing models, and just \$5,800 for the reduced framework. If the reduced framework were used instead of the default ETS implementation, the monetary savings would be \$2,600 for each forecast cycle. Similarly, if the reduced framework were to be used instead of the LightGBM ensemble, financial savings would reach \$4,050 for each forecast cycle. Assuming that forecasts are produced each week, the reduced computational cost translates to monetary savings of about \$135,200 and \$210,600 annually, compared with the other two approaches, respectively. These annual savings would reach \$949,000 and \$1,478,250, respectively, if the forecasts were produced daily.

Note that this example illustrates the minimum possible gains a retailer like Walmart would save using the proposed reduced framework. According to \cite{Seaman2017}, the online marketplace of Walmart (walmart.com) offers 100 million SKUs that may be demanded in 40,000 different ZIP codes. That would form a base of 4,000 billion SKU-ZIP code series. Even assuming just 1\% of these series needs to be forecast regularly, that would suggest 40 times the savings of our previous example (about \$38 million to 59 million in savings in the case of daily forecast cycles). 

\section{Discussion}\label{sec:discussion}

The empirical analysis in Section~\ref{sec:evaluation} as well as the case study in Section~\ref{sec:case_study} show that considering pragmatic subsets of the available forecasting models significantly reduces the required computational time without necessarily negatively affecting the forecast quality in terms of accuracy and uncertainty estimation. Specifically, some software packages suggest avoiding multiplicative trend models for the exponential smoothing family of models. We also show that considering only damped trends (when a trended model is needed) and matching the type of the error component with that of the seasonality component (by excluding models with multiplicative errors and additive trends) brings additional benefits. In the ARIMA family of models, restricting the maximum order of the models also benefits both performance and cost.


On top of the financial savings associated with reduced computational cost, another critical dimension is energy usage and environmental footprint. According to \cite{Naughton2019-vx}, ``the computing power required for today's most-vaunted machine-learning systems has been doubling every 3.4 months.'' Despite the recent trends to use machine (and deep) learning techniques in forecasting, we take one step back and argue for the need for fast, cost-effective solutions. 

To measure the environmental impact of full versus reduced pools of models, we replicated our empirical results based on ETS and as reported in Subsection \ref{sec:performance_cost} on a cloud-based machine provided by Google (GCloud
Windows Server 2012 R2 Datacenter
8 vCPU, 8 GB memory). We recorded the time needed to run the full versus the reduced pool of models and the CO2 emissions as reported by Google.\footnote{For a detailed explanation of how Google reports CO2 emissions, please refer to https://cloud.google.com/carbon-footprint/docs/methodology?hl=el Accessed: 08-Apr-2022} For the 50,045 series considered in Subsection \ref{sec:performance_cost}, the carbon footprint reduction is 0.371 kg CO2e. Scaling this up to the number of forecasts required by a large retailer (as per our example in Section \ref{sec:case_study}; 40 billion SKU-ZIP code combinations) and daily forecast cycles, the yearly carbon footprint reduction is equal to 108,286 tonnes CO2e, equals the emissions produced by 89,000 cars a year\footnote{See also: https://www.acea.auto/figure/average-co2-emissions-from-new-passenger-cars-by-eu-country/ Accessed: 08-Apr-2022} or the annual amount of CO2 absorbed by 3.2 million trees.\footnote{Computations made assuming that a single tree involves 200 kg of biomass and that each kg of such biomass absorbs 0.166 kg of CO2 per year \citep{DOUKAS2021102815}}


\section{Conclusions}\label{sec:conclusions}

Our results are relevant to forecasting research. Over the past 20 years, we have observed a trend to increase the number and complexity of the models in forecasters' toolboxes. Against this trend, our study builds on the arguments of suboptimal \citep{Nikolopoulos2018-wa, Ashouri2019-ac} and simple \citep{Green2015-gf} forecasting solutions.


We empirically tested the performance of small and computationally-efficient forecasting model pools against larger pools on almost 100,000 real-life time series. We noticed that small pools of exponential smoothing or ARIMA models produce results in a fraction of the time larger pools require. Crucially, reduced computational time did not equate to inferior forecasting performance. The reduced exponential smoothing pool led to, on average, more accurate forecasts and better estimation of the forecast uncertainty through prediction intervals. At the same time, low-order ARIMA models offered competitive performance compared with larger models; however, very small models ($K=1$) may not always be the best choice.

We showed that the use of reduced pools of models could be associated with significant savings when one must forecast a vast number of series. We translated the reduced computational cost into monetary savings in the context of large retailers. We suggest that future studies should consider the \textit{cost of the forecast} as an additional performance indicator on top of the traditional forecast error measures. Finally, because this study focused separately on the exponential smoothing and ARIMA families of models, a direction for future research could be the generalization of the findings to other (or even combined) families of forecasting models.

Although we make a case for the simplicity of forecasting methods in our paper, one could argue that Moore's law works against our case. If computing power becomes less and less expensive, how important is it for retailers to consider it in their methods? Recent empirical evidence, however, seems to suggest Moore's Law has slowed considerably. ``The winding down of Moore’s Law means that the technological scaling that drove these historical declines and implicitly underlie the most optimistic assumptions about the spread of ubiquitous computing in the future may no longer hold. Both cost and energy use now seem more likely to increase in lockstep with the scale of cloud computing in the future. Unless there are continuing, significant improvements in software technology, computing costs—and energy use per computation—are unlikely to decline, or even stay constant as computing capacity increases, as was true in the past'' \cite[p. 54]{Flamm2020}. In that context, fast and frugal forecasting may be the way forward for large-scale retail organizations.

Forecasting has always struggled to balance complexity, accuracy, computational time, and interpretability. Modern data science has tipped public opinion toward complexity; we hope that paper serves as Occam's razor and offers a valuable counter perspective. Simple model pools perform incredibly well at a fraction of the computational costs, making forecasting faster, less costly, and more environmentally friendly. Maybe these insights will tip the scales back ever so slightly.


\singlespacing

\renewcommand*{\bibfont}{\normalsize}
\bibliographystyle{elsarticle-harv}
\bibliography{refaggr}

\end{document}